# Electrical properties of Bi-implanted amorphous chalcogenide films

Yanina G. Fedorenko*

Advanced Technology Institute, Department of Electronic Engineering, University of Surrey, Guildford, GU2 7XH, United Kingdom

**Abstract**

The impact of Bi implantation on the conductivity and the thermopower of GeTe, Ge-Sb-Te, and Ga-La-S films is investigated. The enhanced conductivity appears to be notably sensitive to a dose of an implant. Incorporation of Bi in amorphous chalcogenide films at doses up to $1 \times 10^{15}$ cm$^{-2}$ is seen not to change the majority carrier type and activation energy for the conduction process. Higher implantation doses may reverse the majority carrier type in the studied films. Electron conductivity was observed in GeTe films implanted with Bi at a dose of $2 \times 10^{16}$ cm$^{-2}$. These studies indicate that native coordination defects present in amorphous chalcogenide semiconductors can be deactivated by means of ion implantation. A substantial density of implantation-induced traps in the studied films and their interfaces with silicon is inferred from analysis of the space-charge limited current and capacitance-voltage characteristics taken on Au/amorphous chalcogenide/Si structures.

*present address:
Stephenson Institute for Renewable Energy, University of Liverpool, Chadwick Building, Peach St., Liverpool L69 7ZF, United Kingdom

# 1. Introduction

Chalcogenide glasses attract significant attention, because of their potential for use in various solid-state devices including thin-film transistors, neuromorphic information processing, light-emitting diodes, phase-change memory, to name a few [1-3]. One important challenge pertaining to amorphous chalcogenide glasses is low value of electrical conductivity, which could represent a serious limit for their device applications. The majority of chalcogenide glasses are covalently bonded p-type semiconductors. The structural flexibility of the amorphous network allows for the formation of energetically favourable coordination defects, defined as valence-alternation states [4-6]. The valence requirements of an atom introduced into an amorphous chalcogenide semiconductor are typically satisfied, thus preventing the formation of a charged donor or acceptor. To achieve enhanced electron conductivity, amorphous chalcogenide semiconductors have been chemically modified through alloying with transition metals, bismuth and lead in melted glasses [7, 8]. With respect to doping, only bismuth and lead have been found to reverse the majority carrier type, from p- to n-type, in amorphous germanium chalcogenide semiconductors by means of equilibrium doping. Although dopants other than Bi and Pb were not found to result in the type carrier reversal in amorphous chalcogenides, formation of charged point defects, $D^-$ centers, has been detected in Ga-doped GeSe and correlated with smaller $E_A$ values [9].

The origin of the n-type conductivity in bismuth-modified bulk chalcogenide glasses has been related to the high polarizability of bismuth that favours the formation of partially ionic Bi-chalcogen bonds [10]. Strong polarization of Bi or Pb may stabilize the lone-pair electrons of a chalcogen decreasing the energy for the defect pair formation. The latter has been suggested to be approximately equal to $E_A$, the activation barrier for dc conductivity [11]. The higher concentration of the dopant implies the larger stabilization energy which is restricted at the limit by the Coulombic repulsion between the negatively charged defects. Consequently, $E_A$, an indicative of mutual spatial location of chalcogen atoms, has been shown to decrease and saturate as the dopant concentration increases. Assuming the dc conduction is supported by a defect-assisted mechanism, the ratio of positively to negatively charged defects which act as the charge carriers determines the conductivity type. Alternatively, conduction supported by n-type nanoclusters in heterogeneous amorphous chalcogenides has been invoked to explain electron conductivity [12, 13]. These works propose that the *8-N* rule, formulated by Mott [4], does not satisfactorily describe impurity states in the band gap of amorphous chalcogenides whose alloy compositions are substantially heterogeneous. The Bi-related electron conductivity in amorphous chalcogenides has been suggested to be linked to near IR photoluminescence (PL) in various bismuth doped glasses [14]. Given that the Bi ion in melt is polyvalent, and the Bi oxidation state is dependent on the melting temperature of a glass it is not straightforward to conclude on the exact type of defects responsible for the near IR PL. The origin of Bi-related optical centers may involve not only anion vacancies as in crystals, but also other defects of the glass network, thereby modifying defect centers and shifting the luminescence bands in energy [15].

Concerning doping in melt, a slow liquid quench enables a balanced equilibrium with recrystallization acting as a balancing force, thus implying higher doping concentrations are required to shift a melted alloy to a non-homogeneous state of sufficiently perturbed local coordination around an introduced dopant. Non-equilibrium synthesis methods such as thermal evaporation on an unheated substrate have been noticed to facilitated an impurity conductivity with an activation energy $E_A$, smaller than half of the band gap, as in the case of Bi-doped GeTe [16]. The unbalanced charge equilibrium of positively and negatively charged defects would allow to unpin the Fermi level and give rise to charge carriers in the localized band tail states. Hence, non-equilibrium growth or doping techniques are thought to be beneficial in introducing alloy disorder, and driving an alloy into a metastable state. The extremely short time scale within which an ion-implantation cascade interacts with a host solid matrix implies that the structural transformations in the host proceed far from thermodynamic equilibrium. This allows one to consider ion implantation as a fast thermal quench which might enable structural modification of an amorphous matrix. Specific of ion implantation in amorphous chalcogenide semiconductors can be explained taking into account the long diffusion times reaching hundredth nanoseconds at temperature closer to the melting point. As it has been discussed in Ref.17, such values imply slower nucleation and crystal growth rate than that observed in conventional semiconductors and suggest the collective rearrangement of the amorphous network is the main factor determining the crystallization of these materials. Here, the impact of ion implantation, in particular of bismuth, on the electrical properties of lone-pair semiconductors including Ge-Sb-Te, GeTe, and Ga-La-S is considered. It has been anticipated that a slower recovery from the damage during the relaxation process of collision cascades is expected in solids with covalent bonding [18]. Therefore, ion-beam-induced defect accumulation in more ionic Ga-La-S [19] may be expected to proceed differently than in predominantly covalent Ge-Sb-Te and GeTe.

## 2. Experimental

100 nm thick Ge-Sb-Te, GeTe, and Ga-La-S films were prepared by rf sputtering, either on silicon to form heterojunction devices, or on fused silica substrates to probe the conductivity type by means of thermopower measurements. Ion implantation of bismuth was done in a Danfysik ion implanter with ion doses in the range of $1 \times 10^{14}$-$2 \times 10^{16}$ cm$^{-2}$ at energy of 190 keV resulting in depth distributions of Bi which has been exemplified in Fig. 4 of Ref. 14. The shape of the profile could be explained on the basis of the collision processes during implantation without diffusion. Compositional characterization of Ga-La-S film undertaken previously confirmed that the films were $Ga_{26}La_{12}S_{45}O_{17}$. GeTe films were sputtered from a target of a composition equal to 1:1. The as-deposited GeTe films were kept at ambient laboratory conditions and were found to contain substantial amount of oxygen, with an O/Ge molar ratio of 0.8. Upon implantation of bismuth, the GeTe films revealed preferential removal of Te atoms and an increased oxygen content. Oxygen content in Ga-La-S films did not change upon incorporation of Bi as it was verified by means of Rutherford backscattering (RBS) in normal incidence using a 2MV Tandetron accelerator [20]. The Ge-Sb-Te films were prepared by the rf sputtering of a $Ge_2Sb_2Te_5$ target. The samples used in this work were not subjected to anneals.

Samples for current-voltage (*IV*) and capacitance-voltage (*CV*) measurements were fabricated by depositing gold electrodes of area $(1.0–1.2) \times 10^{-2}$ cm$^2$ onto the chalcogenide film surface. *IV* and *CV* traces were recorded using an Agilent B2902A source-measurement unit, a HP4275A LCR meter, or Keithley 4200 SCS. Samples were mounted in an Oxford OptistatDN cryostat connected to a temperature controller to enable such measurements between 77 K and 300 K. The total interface trap density $N_{it}$ (integrated across the Si band gap) was determined from the absolute difference of the flat-band voltage $V_{fb}$ inferred from high-frequency *CV* curves measured on p- and n-type Si MOS-capacitors at 77 K, as described elsewhere [21]. For the n-type and the p-type silicon substrate doping levels (~$2 \times 10^{15}$ cm$^{-3}$ and $5 \times 10^{15}$ cm$^{-3}$ respectively), the difference in $V_{fb}$ obtained from the 77 K *CV* curves almost corresponds to a shift in Fermi level of 1.16 eV. The total density of interface traps, $N_{it}$, can be expressed as $V_{fbn}$ (77 K) $- V_{fbp}$ (77 K) $= E_{gSi}$(77 K) $+ qN_{it}/C_{ox}$, where $E_g$(Si) = 1.16 eV at 77 K, $q$ is the elemental charge, and $C_{ox}$ is the capacitance which accounts for the charge accumulated in a film deposited on a semiconductor substrate.

The relative permittivity of the Ge-Sb-Te and Ga-La-S films needed for the $N_{it}$ estimation and analysis of space-charge-limited conduction in the amorphous chalcogenide films was calculated as $\varepsilon = n^2 - k^2$, where $n$ and $k$ are, respectively, the refractive index and extinction coefficient determined from spectroscopic ellipsometry data. The latter were obtained using a Sopra GES-5 instrument. The measured spectra were fitted with the help of the Tauc-Lorentz dispersion model, in which the Lorentz oscillator term and the Tauc joint density of states describe the optical absorption below and above the optical band gap, respectively [22]. The $\varepsilon$ values corresponding to the optical constants at 830 nm are shown in Table 1. For the amorphous as-deposited Ge-Sb-Te films, the $n$ and $k$ values are taken from the data presented in Figure 5 of Ref. 23. The optical-absorption spectra were obtained on a Varian Cary 5000 UV-vis–near-IR spectrophotometer. The optical-absorption edge was determined as the intersection of straight lines through the baseline in the transparent region and through the absorption edge. Alternatively, an extrapolation of the absorption spectrum in a plot of $(\alpha h\nu)^{1/2}$ vs. $h\nu$ to $\alpha = 0$ gives values of $E_{g\ opt}$. For the Seebeck measurements, the films were deposited on highly resistive fused-silica substrates. An air gap between a copper block and a resistively heated suspended thermal contact provided the temperature gradient between the electrodes. The samples were shielded from external voltages and electrically isolated from ground. Raman spectra were recorded in a backscattering geometry at room temperature using a Renishaw 2000 instrument. The samples were excited either at 514 nm or at 782 nm using a beam focused to 3 μm. The incident laser power was adjusted to ~4 mW to minimize heating effects during the spectrum acquisition time of 150-300 s.

## 3. Results

Fig. 1 exemplifies the temperature dependence of the resistivity and the Seebeck coefficient for Ge-Sb-Te and GeTe films implanted with bismuth. The temperature dependence of the resistivity for Ge-Sb-Te, Figure 1(a), shows two distinctly different activation energies, $E_A$, below and above ~365 K for the case of the as-deposited samples (□) and the samples implanted at a relatively low dose of $5 \times 10^{14}$ ions.cm$^{-2}$ (○). The activation energy determined at temperatures below ~365 K is seen to be insensitive to the implantation dose and falls within the range of $0.35 \pm 0.05$ eV for as-deposited and the Bi-implanted samples. The temperature dependence of the resistivity for bismuth-implanted GeTe films is shown in Fig. 1(b). Compared to the Ge-Sb-Te case, only single activation energy for the conductivity is observed for each sample. At lower temperature, the resistivity values of the unimplanted and lowest bismuth dose samples [c.f. (◁) and (▷) in Fig. 1(b)] are seen to level off, indicating the change in the conduction mechanism. Though single polaron hopping conduction has been suggested for Bi-doped GeSe [24] the conductivity in undoped amorphous chalcogenide materials revealed a correlated barrier hopping of bipolarons between charged defects $D^+$ and $D^-$ in intermediate to high temperature range [25, 26]. In general, it is known that a large temperature dependence of ac conductivity can only be observed in materials which can be characterized by a small negative effective correlation energy and/or a large energy difference between extended states and the Fermi level. A study on the conductivity mechanism in Bi-implanted GeTe films is ongoing and will be presented elsewhere.

The $E_A$ values of 0.7-0.8 eV, characteristic of the unimplanted GeTe films, sharply decrease to 0.2-0.25 eV after implantation of Bi at doses of $5 \times 10^{15}$ ions.cm$^{-2}$ and $2 \times 10^{16}$ ions.cm$^{-2}$ [cf. ($\otimes$) and ($\star$) in Fig. 1(b)]. The dependence of the Seebeck coefficient, $S$, on the bismuth dose is shown in Fig. 1(c) and (d). Qualitatively similar dependences of $S$ on the bismuth dose are observed in both the Ge-Sb-Te and GeTe films, suggesting that the same processes are responsible for the observed modification in the conductivity and in the carrier-transport mechanism. However, the resistivity of GeTe as a function of the bismuth-implantation dose decreases monotonically, whereas for the Ge-Sb-Te films, a non-monotonic dependence of the resistivity is observed, with a minimum value at around $1 \times 10^{15}$ cm$^{-2}$ [cf. ($\triangle$) in Fig. 1(a)]. In the latter case, the use of high implantation doses ($2 \times 10^{16}$ ions.cm$^{-2}$) results in almost a complete removal of the films due to sputtering induced by the impinging ions, thus limiting the maximum bismuth doping level. The increase in the resistivity of the Ge-Sb-Te films may therefore be related to this partial film removal at bismuth implantation doses higher than $1 \times 10^{15}$ cm$^{-2}$ and/or chemical-bond rearrangements. A study of the electrical properties of Ge-Se-Te glasses doped with bismuth in the melt proposed that the decrease in the resistivity could be linked to the formation of Te-Se bonds, and onset of n-type conductivity has been thought to occur as a result of disruption of bonds between the two chalcogen atoms [27]. Both ion-implanted and melt-quenched Ge-Sb-Te and GeTe have been reported to form more Ge-Te bonds compared to as-deposited films [28]. In the case of GeTe, a negative sign for the majority carriers follows from the thermopower data for the film implanted with bismuth at a dose of $2 \times 10^{16}$ ions.cm$^{-2}$, as shown by ($\bigstar$) in Fig. 1(d). The reduction of the optical band gap in the case of Bi implantation in Ga-La-S and Ge-Sb-Te is accompanied by an increase of the relative permittivity (Table 1), suggesting the formation of Bi-chalcogen bonds. These results imply that a chemical interaction between ion-implanted metals and the host amorphous is likely to occur. A decrease in the Seebeck coefficient of GeTe films was also observed upon implantation of Xe or Al at a dose of $3\times10^{16}$ cm$^{-2}$, Fig. 2, indicating that neither the charge state nor the atomic mass of an implanted element are prerequisite requirements to suppress the intrinsic coordination defects in GeTe and diminish the hole conductivity. However, electron conductivity in GeTe was observed only upon incorporation of bismuth.

Figure 3 displays Raman spectra taken on amorphous GeTe films before and after ion-irradiation with Bi. The as-deposited amorphous GeTe, Figure 3(a), reveals peaks at 125 cm$^{-1}$, 137 cm$^{-1}$, 155 cm$^{-1}$, 180 cm$^{-1}$, 217 cm$^{-1}$, and 266 cm$^{-1}$, in good agreement with earlier observations [17], [29, 30, 31]. Depending on local surrounding of Ge, the Raman-active modes in a-GeTe have been interpreted differently. When Ge atoms are considered to be both in defective octahedral sites and tetrahedral sites the bands in the interval below 190 cm$^{-1}$ have been ascribed to vibrations of atoms in defective octahedral units, and the Ge tetrahedra have been suggested to contribute to the Raman spectrum above 190 cm$^{-1}$ [30]. Taking into account possible presence of Ge in octahedral structural units the bands at 125 cm$^{-1}$ and 166 cm$^{-1}$ might be associated with corner- and edge-sharing octahedral units. The Raman band at 217 cm$^{-1}$ is likely to originate from antisymmetric stretching modes of Ge-Ge bonds, with the frequency being higher for the larger number of Ge-Ge bonds in a tetrahedral unit. Further, a broad and weak band at 275 cm$^{-1}$ has been attributed to Ge-Ge vibrations in Ge-rich a-GeTe, possibly indicating segregation of Ge. Here, a band at 266 cm$^{-1}$ could be ascribed to Ge-Ge bonds. A shift towards lower frequencies might indicate a weaker bond. As our samples contained oxygen, a band centred at 137 cm$^{-1}$ could be due to ageing of the GeTe samples, similarly to the band observed at 139 cm$^{-1}$ in oxidized amorphous GeTe [17]. Figure 3(b) shows the Raman spectrum following ion implantation with bismuth at a dose of $2\times10^{16}$ ions.cm$^{-2}$. The spectrum is dominated by a peak at 155 cm$^{-1}$. This band, coupled with that at 125 cm$^{-1}$, would suggest a contribution from defective octahedral Ge sites in a-GeTe. However, a band at 125 cm$^{-1}$ is not resolved in Bi-implanted GeTe hinting that ion implantation removes the weakest bonds. Instead, a band at 166 cm$^{-1}$ appears which can be assigned to edge-sharing octahedral units. The bands observed in the as-deposited GeTe at frequencies 217 cm$^{-1}$ and 266 cm$^{-1}$ are likely to shift to 199 cm$^{-1}$ and 253 cm$^{-1}$ in the bismuth-implanted sample. A peak at 133 cm$^{-1}$ is blue-shifted to a moderate extent in respect to the position in the unimplanted GeTe. Based on the results it is plausible to suggest that tetrahedral structural units in GeTe become distorted, and Ge atoms might partially occupy defective octahedral sites. This observation corroborates the results on structural re-arrangement in Ge-implanted GeTe and the observed reduced fraction of the tetrahedral species, during the spike heating in ion implantation, in a way to facilitate the formation of GeTe6 octahedra, energetically less favourable structural unites of the rocksalt crystalline GeTe. Ion implantation in Ge-Sb-Te could be expected to result in similar structural modifications as the structure of amorphous Ge-Sb-Te may contain Ge in both tetrahedral and octahedral configuration [32]. Furthermore, as far as the tetrahedral coordination of Ge is favored by homopolar Ge-Ge bonds, doping of amorphous GeTe by ion implantation of Bi is seen to decrease the quantity of Ge-Ge bonds and Ge-tetrahedra. Likewise, damping of in-phase breathing vibration modes extended along GeSe4 tetrahedral chain units in GeSe which form a band in Raman spectra at around 200 cm$^{-1}$ occurred in Bi-implanted GeSe samples (not shown here). The observations are consistent with formation of a more chemically ordered state in an amorphous chalcogenide films occurring simultaneously with an increase in structural disorder and resulting in delocalization of electronic and vibrational states [33, 34]. Concomitantly, the narrowing of the optical band gap and higher permittivity have been observed in GeSe under high pressure. In equilibrium, the effect of pressure is generally to rise both the melting and boiling points. In non-equilibrium processes of ion implantation the energy dissipation in the thermal spike can be regarded as propagation of a shock wave with hypervelocity when the local cascade volume decompresses and transmits pressure to the surrounding material. The Vineyard's thermal spike model [35] can be used to estimate that the decay from $10^4$ K to ambient temperature is completed within $\sim 10^{-10}$-$10^{-9}$ s. A characteristic particle kinetic energy in a cascade is on the order of a few eV per particle. The solid phases during thermal spike relaxation are formed polymorphically being forced to evolve locally without the aid of long-rage chemical diffusion. This could explain why non-equilibrium doping by means of ion implantation requires much lower dopant concentrations to obtain n-type conductivity in amorphous chalcogenides. Although in the glass transition region structural relaxation rates increase sharply as the glassy state tends to use the opportunity to readjust toward the equilibrium, the pressures momentary imparted on a molten volume can be in a range of a few GPa that promotes crystal nucleation or formation of an amorphous phase which can be characterized by the higher entropy. This is consistent with coexistence of structurally and thermodynamically distinct but otherwise compositionally identical amorphous phases of the alloy.

No resistivity or thermopower data could be determined for the case of as-deposited Ga-La-S films, due to their excessively high resistance. For this reason, the impact of ion implantation on the electrical properties of Ga-La-S films can be assessed using $JV$-characteristics taken on metal-amorphous chalcogenide-silicon structures. Figures 4(a, b) compare the forward $JV$-characteristics for Ge-Sb-Te (a) and Ga-La-S (b) on a *log-log* scale. Space-charge-limited conduction is apparent, as follows from the steeper $JV$-curves and the temperature-induced voltage shift towards higher voltages as the temperature decreases. The data can be fitted by the expression $J \propto V^m$ with $m$ ranging from 2.4 to 4.9 at 300 K for different implanted samples, evidencing an exponential trap distribution ($m > 2$), as is expected for traps originating from surface defects and structural disorder. Quantitative information about the traps can be obtained by extrapolating the $JV$-curves with voltage [36]. If the charge traps are distributed in energy, there will be a gradual filling with an increasing electric field at all temperatures until, at a certain critical voltage $V_c$, all traps are filled. This critical voltage is independent of temperature and is given by $V_c = qN_t d^2/2\varepsilon\varepsilon_0$. The $V_c$ and $N_t$ data obtained by extrapolating the *JV*-curves are listed in Table 1, indicating an apparent increase in the trap density in the Bi-implanted films, and more so in Ge-Sb-Te.

Figures 5(a) and (b) show dispersion-free Mott-Schottky curves taken at 77 K on the heterojunctions (HJs) formed on using p- and n-type silicon substrates. At 77 K, the vast majority of the interface traps are below the Fermi level. This permits a correct determination of the flat-band voltage $V_{fb}$ in the heterojunction devices. Typical *JV*-characteristics of p-n and p-p Ga-La-S/Si(100) and Ge-Sb-Te/Si(100) HJs at 300 K and 77 K are shown in the insets of Figure 5. The rectification behaviour is observed independently of the conductivity type of crystalline silicon and the width of the optical gap ranging from 0.8 eV in amorphous GeTe [37] to 2.4 eV in Ga-La-S [38]. It is plausible to suggest that rectification in amorphous chalcogenide/silicon heterojunctions is determined by the silicon bend bending rather than the band offsets at the amorphous chalcogenide/silicon interface. Qualitatively similar rectifying behaviour of *IV*-curves has been observed in $Eu_xGd_{1-x}O/Si$ and EuO/Si HJs [39] with an explanation pointing out towards the band bending near a ferromagnetic film surface due to negative charge present in the films.

Figure 6 compares 100 kHz *CV*-curves of both n-type and p-type amorphous chalcogenide/silicon HJs at 300 and 77 K. Upon cooling to 77 K, the *CV* curves of the p-type samples shift to a negative voltage, indicating the presence of donor-type traps in the lower half of the band gap. The *CV* curve of the n-type samples shifts to a more positive gate voltage indicating that acceptor-type traps are present above the Si midgap energy. The $N_{it}$ and the $V_{fb}$ values determined from the 100 kHz *CV* curves at 77 K shown in Table 1 are distinctly different for HJs formed on p- and n-type silicon, implying an asymmetry in the energy distribution of the interface states with respect to the silicon midgap energy. In the unimplanted Ge-Sb-Te/Si and Ga-La-S/Si devices, values of $V_{fb}$ observed for the n-type HJs are larger than that for HJs on p-type silicon. This implies a somewhat higher density of interface traps in the upper part of the silicon bandgap. Implantation of bismuth in Ga-La-S and Ge-Sb-Te up to a dose of $1x10^{15}$ ions.cm$^{-2}$ does not change this tendency.

Implantation of Ga-La-S at a higher bismuth dose of $3\times10^{15}$ ions.cm$^{-2}$ introduces a substantial shift of $V_{fb}$ towards more negative voltages in the case of HJs on p-type silicon. This could be interpreted as evidence for interface trap generation in the lower half of the silicon bandgap. However, the 77 K-*CV* traces recorded on implanted Ge-Sb-Te/Si and Ga-La-S heterojunctions, Fig. 6, do not reach a capacitance which corresponds to depletion and accumulation of silicon by majority carriers, thus precluding a correct determination of $V_{fb}$ and $N_{it}$. While the *CV* traces taken at 300 K and 77 K show a considerable Gray-Brown shift, indicating a high interface trap density in the silicon bandgap, the hysteresis of both 300 K and 77 K *CV* curves is negligibly small, suggesting that the studied chalcogenide/silicon interfaces should not suffer from significant charge instabilities. A deep-depletion phenomenon is observed at 77 K, as depicted by the dotted-line *CV* curve in Figure 6, analogous to ultrathin-gate oxide MOS-devices when a direct tunnelling mechanism dominates the current behaviour, and the carrier rate through the oxide is rather large. Minority carriers generated in the substrate in this case directly tunnel through the oxide under an excess applied gate bias. To preserve electrical neutrality, the depletion region expands beyond that in thermal equilibrium, and the total capacitance continues to decrease below its thermal-equilibrium value. Another prominent feature observed here, the roll-off of the capacitance at the silicon band bending corresponding to the accumulation of the majority carriers, is a common feature observed in tunnel-oxide MOS-device behaviour. Although the roll-off capacitive behaviour of tunnel-oxide MOS-devices appears as a combined effect of direct charge-carrier tunnelling and the increased surface roughness [40], neither of these two factors are determining in the case of 100 nm thick amorphous chalcogenide films. The charge carriers are therefore likely to traverse the films by means of trap-assisted tunnelling. It is apparent that ion implantation, though to a different extent, introduces bulk traps in the amorphous chalcogenide films and their interfaces with silicon. The latter could indicate presence of hydrogen or strained bonds in the as-deposited films.

Non-equilibrium specific of ion implantation implies that the amount of energy released within the collision cascades can be sufficient to lead to a "displacement-spike" effect, particularly when implanting heavy elements. The recovery of the damage is an activated process of collective atomic re-arrangement; the energy barriers for damage recovery actually reflect the probability for re-crystallization which is dependent on the energy density within a cascade. It is then plausible to suggest that the likely effect of ion implantation is to induce structural disorder which increases number of overlapping of electron orbitals and causes broadening of the energy bands. In particular case of bismuth and lead dopants, a moderate electronegativity difference (~0,5) with a chalcogen atom may allow formation of a polar covalent bond. As it has been supposed [11], hybridization of Bi and Pb can be sp3d2, and the energy position of the sp3d2 band can be such that the sp3d2 levels are higher in energy than the levels of the lone-pair electrons of the negatively charged chalcogen atom. Similar to the case of Pb in GeSe, Bi in GeTe may weaken Ge-Te bonds upon doping, widen the antibonding σ*-band, and lower the energy levels.

## 4. Conclusions

To conclude, ion implantation of bismuth in lone-pair amorphous chalcogenide semiconductors of dissimilar chemical compositions and bonding types modified electrical properties of the films and enhanced the conductivity in a qualitatively similar manner. The enhanced conductivity comes as a result of ion beam induced defect accumulation in the films as inferred from the analysis of space-charge limited conduction. It can be speculated that the dose dependent kinetics of defect formation is sensitive to the uniformity of the bond distribution within as-grown films that could explain dissimilarities in the ion-beam induced modulation of the conductivity in each particular amorphous chalcogenide semiconductor. The carrier type reversal in GeTe is accompanied by a significant broadening of the Raman spectra to the extent that the vibrational modes ascribed to Ge-Ge bonds become indistinguishable. In agreement with the previously observed reduction of tetrahedral structural units upon Ge implantation in GeTe, our observations imply that implantation of Bi in GeTe could facilitate formation of a more chemically ordered state of the alloy. In agreement with increase in overall coordination of the amorphous network, this explains why doping of amorphous chalcogenide films becomes feasible.

In a view of a practical perspective, ion implantation in conjunction with large scale thin film deposition methods such as the atomic layer deposition [41] may certainly be used in technological development of devices employing amorphous chalcogenide films. The ability of ion irradiation to avoid detrimental effects associated with interdiffusion and phase formation that usually accompany thermal processing may be employed in semiconductor technology, for example, to increase grain growth in polycrystalline thin films [42].

**ACKNOWLEDGEMENTS**

The author acknowledges insightful discussions with R. J. Curry, S. R. Elliott, R.M. Gwilliam, K. Homewood, D. W. Hewak, and M. A. Hughes in the course of this work.

This work was funded by EPSRC grants EP/I018417/1, EP/I018050/1 and EP/I019065/1

**Figure captions**

Figure 1. The resistivity $\rho$ of Ge-Sb-Te (a) and GeTe films (b) represented by open symbols and the Seebeck coefficient, $S$, of Ge-Sb-Te (c) and GeTe (d) represented by filled symbols for unimplanted (■,□,◀,◁) and samples implanted at varying bismuth doses of $5\times10^{13}$ ions.cm$^{-2}$ (✳), $1\times10^{14}$ ions.cm$^{-2}$ (▶,▷), $5\times10^{14}$ ions.cm$^{-2}$ (●,○), $1\times10^{15}$ ions.cm$^{-2}$ (▲,△,×), $5\times10^{15}$ ions.cm$^{-2}$ (▽,⊗,⊕), $1\times10^{16}$ ions.cm$^{-2}$ (◆,◇), and $2\times10^{16}$ ions.cm$^{-2}$ (★,✧).

Figure 2. The Seebeck coefficient of amorphous GeTe films before (□) and after implantation with Xe at a dose of $3\times10^{16}$ ions.cm$^{-2}$ (○), Al at a dose of $2\times10^{16}$ ions.cm$^{-2}$ (△), and Al at a dose of $3\times10^{16}$ ions.cm$^{-2}$ (▽).

Figure 3. Raman spectra of as-deposited GeTe (a) and after ion implantation with Bi at a dose of $2\times10^{16}$ ions.cm$^{-2}$ (b).

Figure 4. Forward-bias $JV$-characteristics for (a) Ge-Sb-Te and (b) Ga-La-S HJs formed on n-type silicon before implantation (▲,△) and bismuth implantation with doses of: $1\times10^{14}$ ions.cm$^{-2}$ (●,○), $5\times10^{14}$ ions.cm$^{-2}$ (□,■), $1\times10^{15}$ ions.cm$^{-2}$ (◆,◇), $3\times10^{15}$ ions.cm$^{-2}$ (▼,▽). Open and filled symbols correspond to measurements at 300K and 77K, respectively.

Figure 5. Mott-Schottky plots for unimplanted (a) Ge-Sb-Te and (b) Ga-La-S heterojunctions formed on p-type (open symbols) and n-type (filled symbols) silicon measured at frequencies of 100 kHz (○,●), 200 kHz (▲,△), 400 kHz (▼,▽), and 1 MHz (◆,◇) at 77K. The insets show $JV$-curves taken at 300K (filled symbols) and 77K (open symbols).

Figure 6. 100kHz-$CV$ characterisation undertaken at 300 K (○,□) and 77 K (—,--) on the GeTe/Si (a) and the Ga-La-S/Si (b) formed on n-type (□, --) and p-type (○,—) silicon, respectively, after implantation of bismuth at a dose of $1\times10^{16}$ ions.cm$^{-2}$ in Ge-Sb-Te and of $3 \times 10^{15}$ ions.cm$^{-2}$ in Ga-La-S. In (a) and (b) the 77 K data has been multiplied by a factor 10 for clarity.

TABLE I. Carrier doping level in the silicon substrate $N$, the flat-band voltage $V_{fb}$, the critical voltage $V_c$, the total interface trap density $N_{it}$, the trap density $N_t$, and the relative permittivity for Ge-Sb-Te and the Ga-La-S films on silicon, as influenced by implantation of Bi.

| Bi (ions.cm$^{-2}$) | $N$ (cm$^{-3}$) p-Si | $N$ (cm$^{-3}$) n-Si | $V_{fb}$ (V) (77K) n-Si HJ | $V_{fb}$ (V) (77K) p-Si HJ | $V_c$ (V) | $N_t$ (10$^{17}$.cm$^{-3}$) | $N_{it}$ (10$^{12}$.cm$^{-2}$) | $\varepsilon$ |
|---|---|---|---|---|---|---|---|---|
| Ge-Sb-Te | | | | | | | | |
| as-dep. | 3.1x10$^{15}$ | 1.9x10$^{13}$ | 1.2 | -0.3 | 1.27 | 1.7 | 0.1 | 12 |
| 5x10$^{14}$ | 4.2x10$^{15}$ | 3.9x10$^{15}$ | 1.0 | -1.7 | 4.5 | | | |
| 1x10$^{15}$ | 6.4x10$^{15}$ | 1.7x10$^{15}$ | 7.0 | -3.2 | 5.7 | | | |
| 3x10$^{15}$ | 7.0x10$^{15}$ | 7.9x10$^{14}$ | 5.3 | -3.0 | 5.9 | 10.7 | 2.4 | 16.5 |
| Ga-La-S | | | | | | | | |
| as-dep. | 6.4x10$^{15}$ | 1.2x10$^{15}$ | 15.5 | -1.6 | 2.8 | 1.0 | 14.6 | 3.4 |
| 5x10$^{14}$ | 1.7x10$^{16}$ | 2.6x10$^{15}$ | 18.9 | -3.0 | 6.8 | | | |
| 1x10$^{15}$ | 1.4x10$^{16}$ | | 16.9 | -14.3 | 7.6 | | | |
| 3x10$^{15}$ | 1.2x10$^{16}$ | 8.9x10$^{15}$ | 16.5 | -172.3 | 9 | 3.8 | ~10$^2$ | 3.8 |

FIGURE 1. *Electrical properties of Bi-implanted amorphous chalcogenide films*

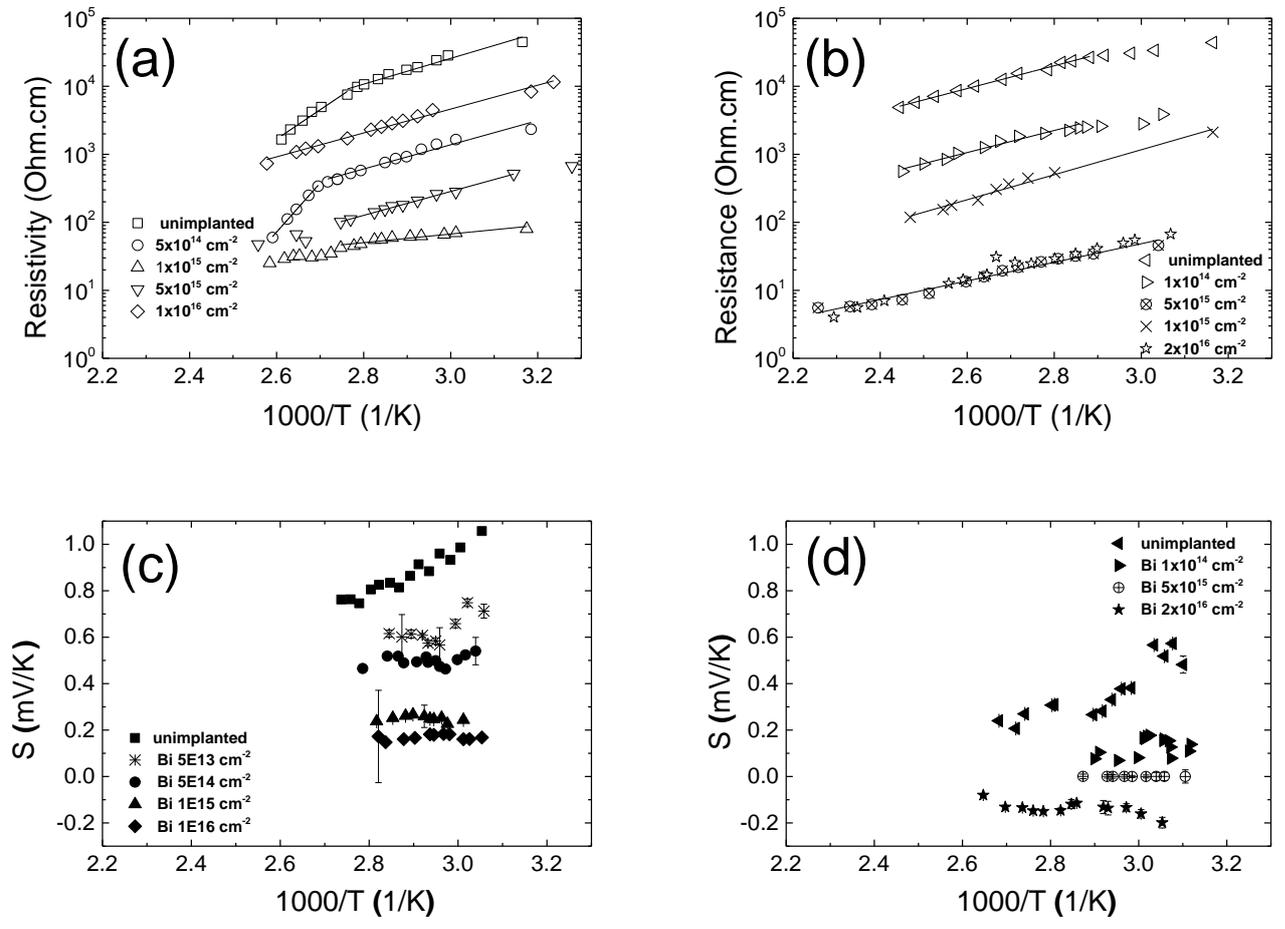

FIGURE 2. *Electrical properties of Bi-implanted amorphous chalcogenide films*

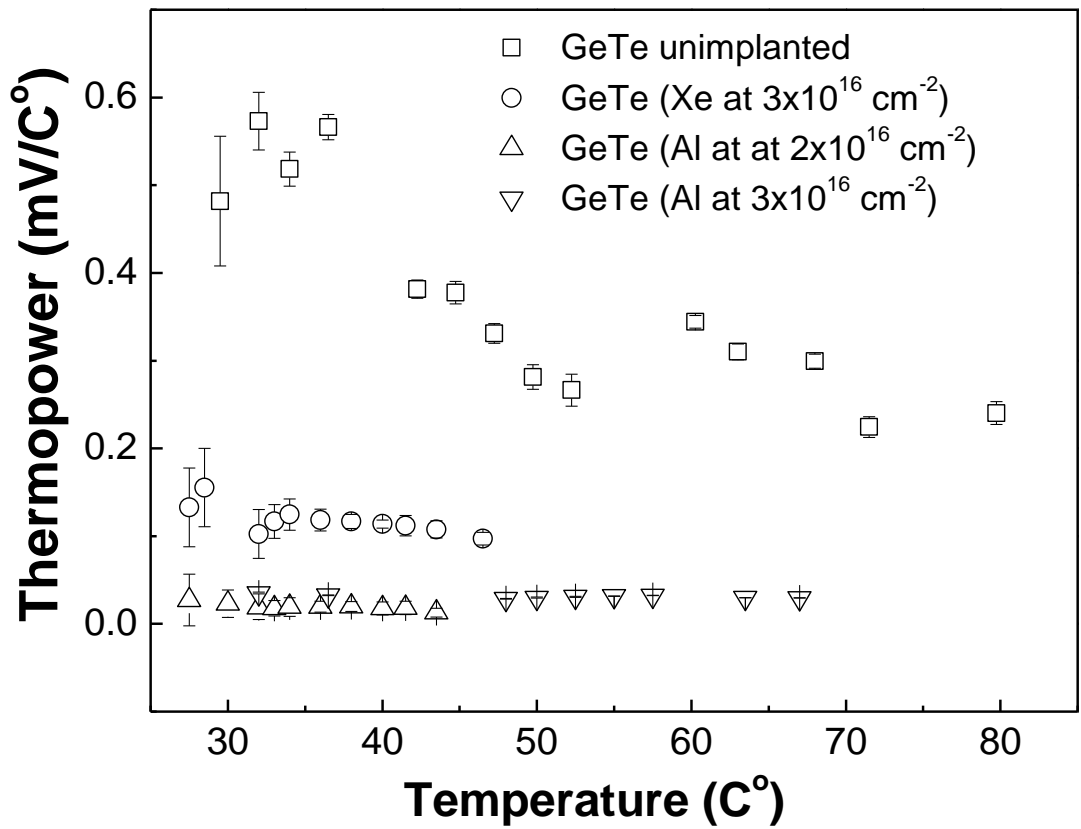

FIGURE 3. *Electrical properties of Bi-implanted amorphous chalcogenide films*

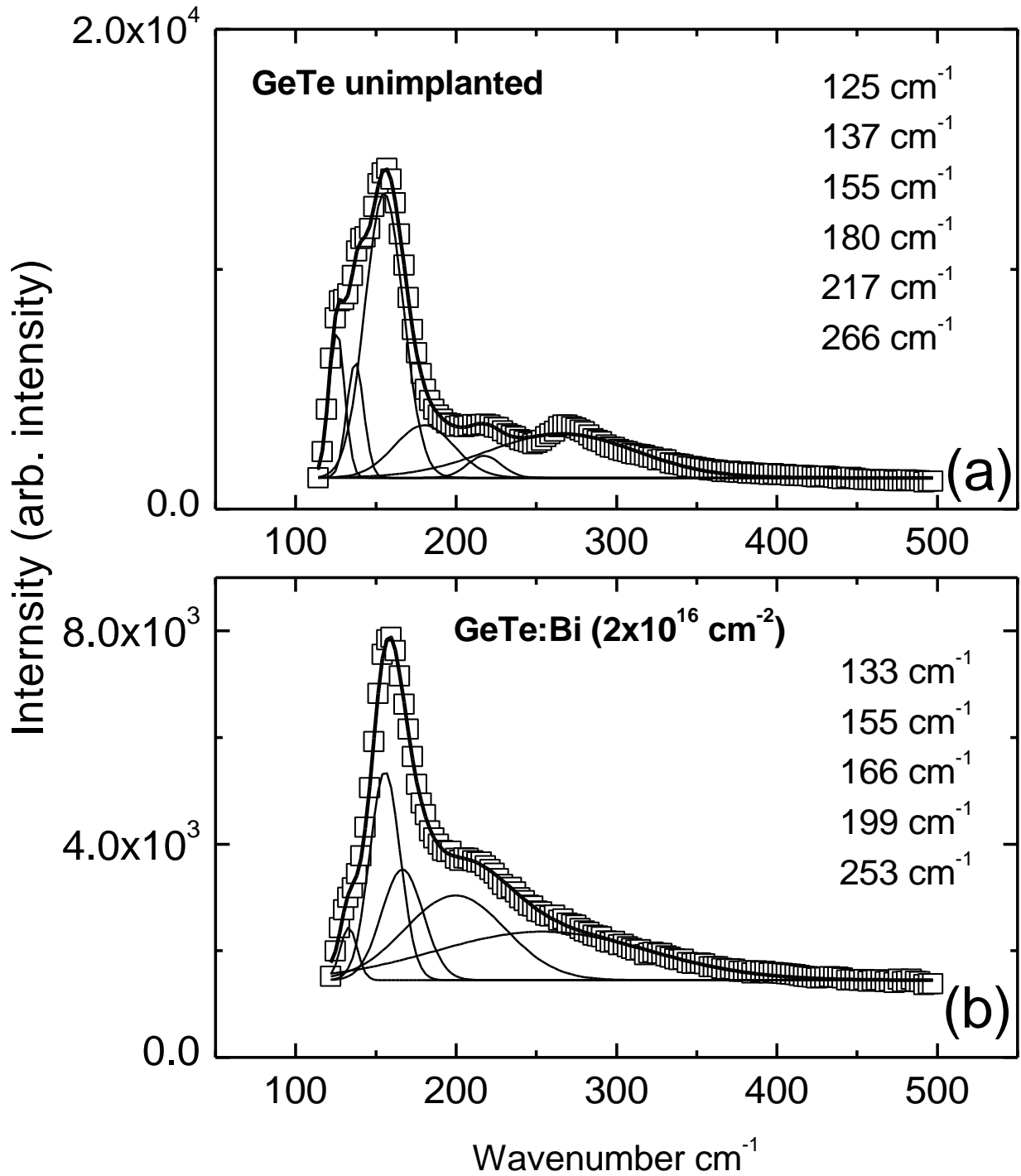

FIGURE 4. *Electrical properties of Bi-implanted amorphous chalcogenide films*

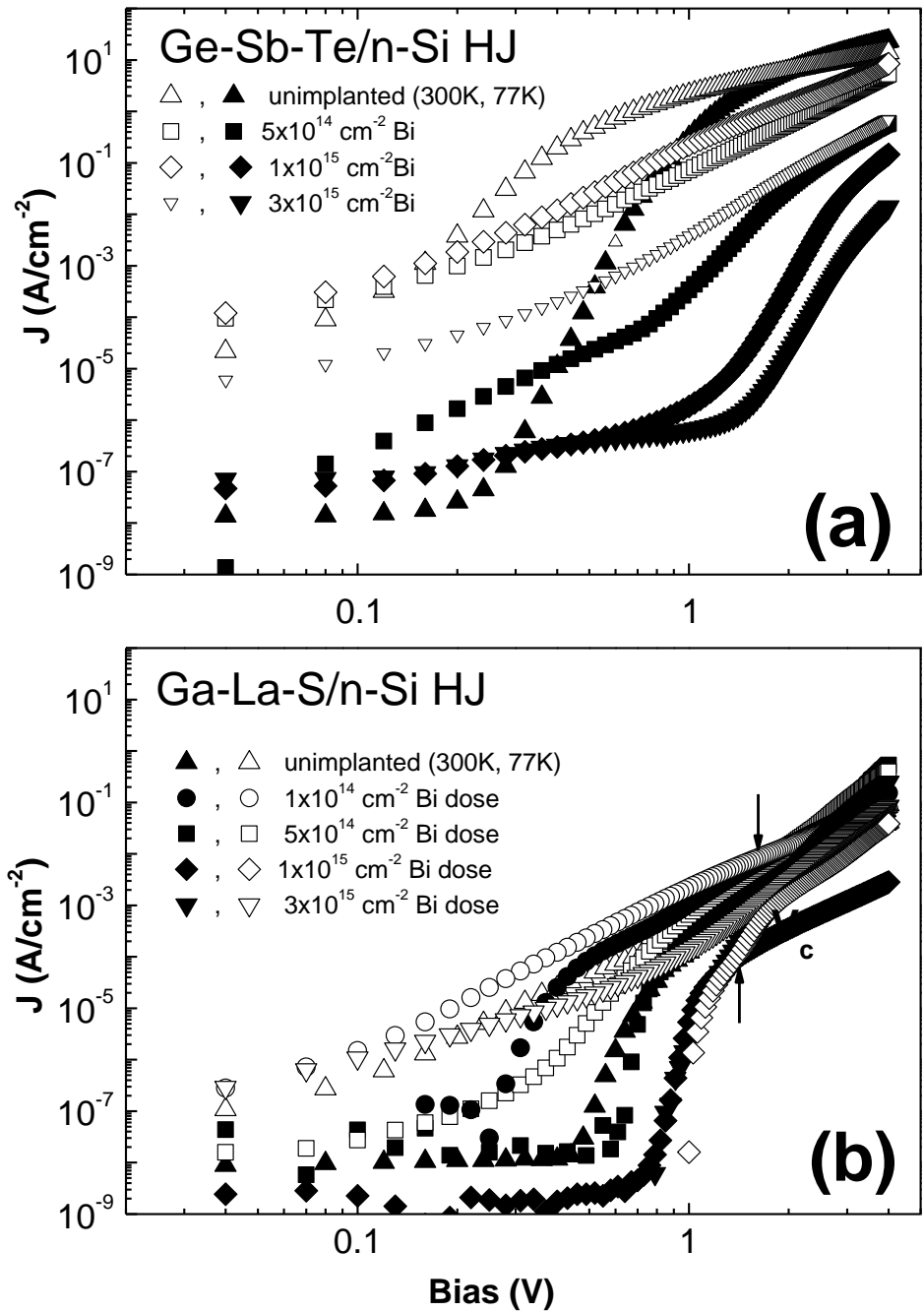

FIGURE 5. *Electrical properties of Bi-implanted amorphous chalcogenide films*

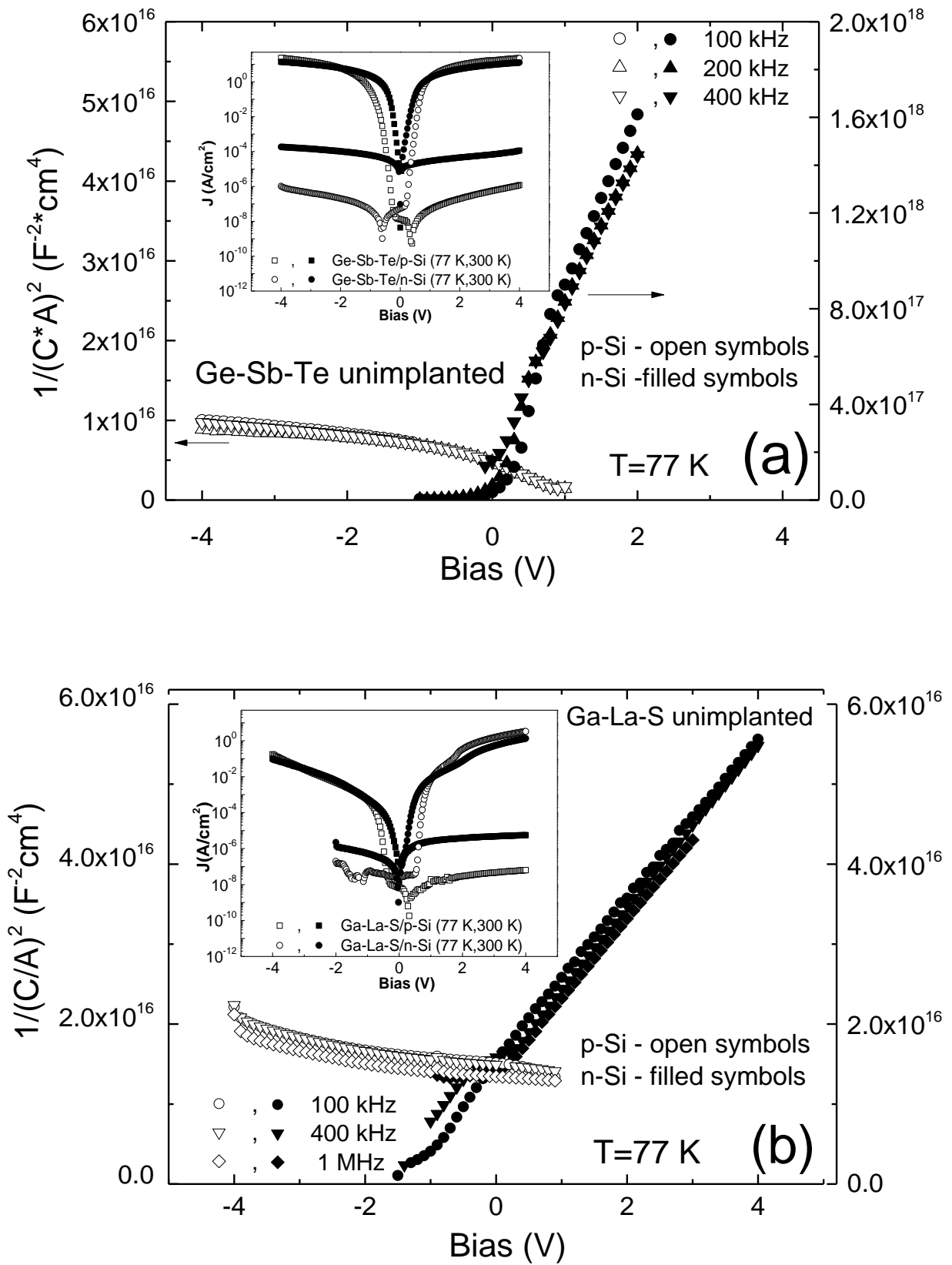

FIGURE 6. *Electrical properties of Bi-implanted amorphous chalcogenide films*

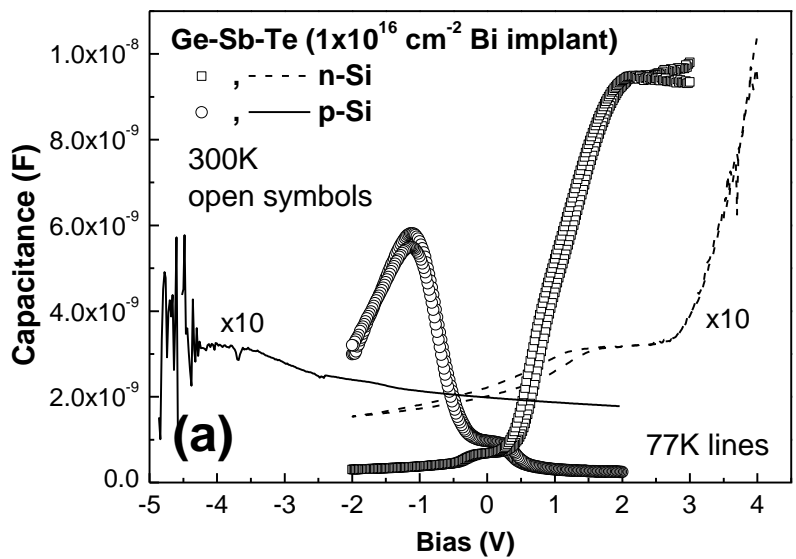

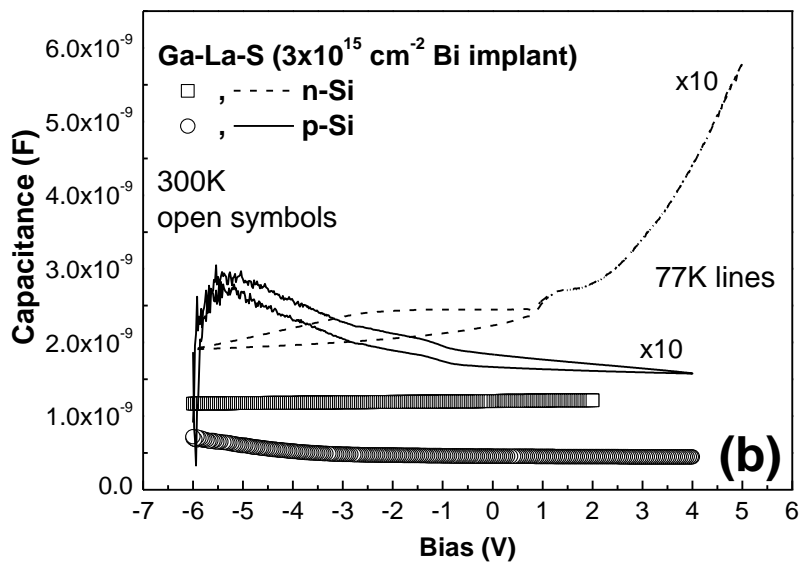